\documentclass[12pt]{article}
\usepackage{amsmath}
\usepackage{graphicx}
\usepackage{enumerate}
\graphicspath{{figures/}}
\usepackage{natbib}
\usepackage{url}

\newcommand{\blind}{1}
\usepackage{amsmath}
\usepackage{bbm}
\usepackage{graphicx,psfrag,epsf}
\usepackage{enumerate}
\usepackage{natbib}
\usepackage{amsfonts}
\usepackage{amsthm}
\usepackage{amssymb}
\usepackage{bm}
\usepackage{caption}
\usepackage{multirow}
\usepackage{booktabs}
\usepackage{url} 
\usepackage{color}
\usepackage[colorlinks,citecolor=blue,urlcolor=blue]{hyperref}

\newcommand{\ve}[1]{{\mbox{\boldmath ${#1}$}}}

\usepackage{xr}
\makeatletter
\newcommand*{\addFileDependency}[1]{
	\typeout{(#1)}
	\@addtofilelist{#1}
	\IfFileExists{#1}{}{\typeout{No file #1.}}
}
\makeatother

\newcommand*{\myexternaldocument}[1]{%
	\externaldocument{#1}%
	\addFileDependency{#1.tex}%
	\addFileDependency{#1.aux}%
}
\myexternaldocument{supp20250420}

	\newtheorem{theorem}{Theorem}
	
	\newtheorem{proposition}{Proposition}
	\newtheorem{definition}{Definition}
	\newtheorem{corollary}{Corollary}

 \newtheorem{condition}{Condition}
 \newtheorem{example}{Example}

\begin{document}
\def\spacingset#1{\renewcommand{\baselinestretch}%
{#1}\small\normalsize} \spacingset{1}
\def\var{{\rm Var}\,}
	\def\blue#1{\textcolor{blue}{#1}}
	\def\red#1{\textcolor{red}{#1}}

\if1\blind
{
  \title{\bf Online differentially private inference in
 stochastic gradient descent}
\author{
Jinhan Xie\textsuperscript{1}\thanks{The first two authors contributed equally to this work.},
Enze Shi\textsuperscript{2}\footnotemark[1],
Bei Jiang\textsuperscript{2}\thanks{Corresponding author},
Linglong Kong\textsuperscript{2},
Xuming He\textsuperscript{3}
}

\date{}
\maketitle
} \fi

\vspace{-1cm}
\begin{center}
\textsuperscript{1}Yunnan Key Laboratory of Statistical Modeling and Data Analysis, Yunnan University\\
\textsuperscript{2}Department of Mathematical and Statistical Sciences, University of Alberta\\
\textsuperscript{3}Department of Statistics and Data Science, Washington University in St. Louis\\

\end{center}

\if0\blind
{
  \bigskip
  \bigskip
  \bigskip
  \begin{center}
    {\LARGE\bf Online differentially private inference in
 stochastic gradient descent}
\end{center}
  \medskip
} \fi

\begin{abstract}
 We propose a general privacy-preserving optimization-based framework for real-time environments without requiring trusted data curators. In particular, we introduce a noisy stochastic gradient descent algorithm for online statistical inference with streaming data under local differential privacy constraints. Unlike existing methods that either disregard privacy protection or require full access to the entire dataset, our proposed algorithm provides rigorous local privacy guarantees for individual-level data. It operates as a one-pass algorithm without re-accessing the historical data, thereby significantly reducing both time and space complexity.  {We also introduce online private statistical inference by conducting two construction procedures of valid private confidence intervals. 
  We formally establish the convergence rates for the proposed estimators and present a functional central limit theorem to show the averaged solution path of these estimators weakly converges to a rescaled Brownian motion,  providing a theoretical foundation for our online inference tool.} Numerical simulation experiments demonstrate the finite-sample performance of our proposed procedure, underscoring its efficacy and reliability. Furthermore, we illustrate our method with an analysis of {two datasets: the ride-sharing data and the US insurance data, showcasing its practical utility}.
\end{abstract}

\noindent%
\vfill

\newpage
\spacingset{1.9}

 \section{Introduction}
\renewcommand{\theequation}{1.\arabic{equation}}
\setcounter{equation}{0}

Online learning involves methods and algorithms designed for making timely inferences and predictions in a real-time environment, where data are collected sequentially rather than being static as in traditional batch learning. By not requiring the simultaneous use of the entire sample pool, online learning alleviates both storage and computation pressures. To date, various statistical methods and algorithms for online estimation and inference have been proposed, including aggregated estimating equations \citep{lin2011aggregated},  cumulative estimating equation \citep{schifano2016online}, renewable estimator \citep{luo2020renewable}, and stochastic gradient descent (SGD) \citep{robbins1951stochastic, polyak1992acceleration,chen2020statistical}. Alongside online learning, another crucial consideration  is the preservation of privacy, especially with the growing availability of streaming data. For example, financial institutions, such as banks and credit card companies, often leverage customer-level data for fraud detection and personalized service offerings, which need to continuously update and refine their models in real-time as new transaction data are collected. This aids not only in monitoring transactions to detect unusual activities but also in designing targeted financial products and promotions (e.g., offering personalized loan rates or credit card rewards based on individual spending patterns); see \cite{maniar2021differential, davidow2023privacy}. However, individual customer data are usually sensitive and irreplaceable, and it is of great importance to prevent the leak of personal information to maintain customer trust and confidence, as well as to help businesses avoid financial losses and reputational damage \citep{hu2022privacy, fainmesser2023digital}.

In response to the increasing demand for data privacy protection, differential privacy (DP) has emerged as a widely adopted concept \citep{dwork2006calibrating}. It has been successfully applied in numerous fields, including healthcare, financial services, retail, and e-commerce \citep{dankar2013practicing, zhong2019commerce, wang2022protection}.  DP procedures ensure that an adversary cannot determine whether a particular subject is included in the dataset with high probability, thereby providing robust protection for personal information. In the literature, two main models have emerged for DP: the central model, where a trusted third party collects and analyzes data before releasing privatized results, and the local model, where data is randomized before sharing. Several works have focused on designing private algorithms to guarantee DP under the trusted central server setting across various applications, including, but not limited to, machine learning, synthetic data generation, and transfer learning; see \cite{karwa2016inference, ponomareva2023dp, li2024federated}. A significant limitation of the standard central model is the assumption of a trusted data curator with access to the entire dataset. However, in scenarios such as smartphone usage, users may not fully trust the server and prefer not to have their personal information directly stored or updated on any remote central server.
In contrast to the central DP, user-level local differential privacy (LDP) does not rely on a trusted data collector, placing the privacy barrier closer to the users, which has seen adoption in practice by
several companies that analyze sensitive user data, including Apple \citep{tang2017privacy}, Google \citep{erlingsson2014rappor}, and Microsoft \citep{ding2017collecting}. Recently, there has been growing interest in studying LDP from a statistical
perspective, including density estimation \citep{sart2023density}, mean and median estimation \citep{duchi2018minimax}, nonparametric estimation problems \citep{rohde2020geometrizing}, and change-point detection \citep{berrett2021locally}.

The aforementioned methods mainly address the inherent trade-off between privacy protection and efficient statistical inference, focusing on identifying what optimal privacy-preserving mechanisms may look like.  The literature on private inference, particularly in quantifying the uncertainty of private estimators, is relatively limited even in offline settings and has recently gained attention in computer science \citep{karwa2018finite, wang2019differentially, chadha2021private}. This issue has also been studied in the statistical literature. \cite{sheffet2017differentially} provided finite-sample analyses of conservative inference for confidence intervals regarding the
coefficients from the least-squares regression. 
\cite{barrientos2019differentially} presented algorithms for comparing the sign and significance level of privately computed regression coefficients that satisfy DP. \cite{avella2021privacy} introduced the M-estimator approach to conduct DP statistical inference using smooth sensitivity. \cite{alabi2023differentially} proposed DP hypothesis tests for the multivariate linear regression model. 
 \cite{avella2021differentially} investigated optimization-based approaches for Gaussian differentially private M-estimators. \cite{zhao2024private} extended this to objective functions without local strong convexity and smoothness assumptions. \cite{zhang2024differentially} considered estimation and inference problems in the federated learning setting under DP. Despite these efforts, a significant gap remains in developing a statistically sound approach for conducting statistical inference, such as constructing confidence intervals based on private estimators, particularly in the online setting, where theoretical results for statistical inference are largely lacking. Recently, \cite{liu2023online} proposed a self-normalizing approach to obtain confidence intervals for population quantiles with privacy concerns. However, this method is specifically tailored to population quantiles.

In this paper, we introduce a rigorous LDP framework for online estimation and inference in optimization-based regression problems with streaming data, addressing the challenge of constructing valid confidence intervals for online private inference. Specifically, we propose a fully online, locally private framework for real-time estimation and inference without requiring trusted data collectors. To the best of our knowledge, this is the first time in the literature on online inference problems where the issue of privacy is systematically investigated. The main contributions of this paper are summarized as follows:
\begin{enumerate}
    \item {To eliminate the need for trusted data collectors}, we design a local differential privacy stochastic gradient descent  algorithm (LDP-SGD) for streaming data that requires only one pass over the data. Unlike prior works, our algorithm provides rigorous individual-level data privacy protection in an online manner. {Meanwhile, we propose two online private statistical inference procedures. One is to conduct a private plug-in estimator of the asymptotic covariance matrix. The other is the random scaling procedure, bypasses direct estimation of the asymptotic variance by constructing an asymptotically pivotal statistic  for valid private inference in real time.}
    \item {From a theoretical standpoint, we establish the convergence rates of the proposed online private estimators. In addition, we prove a functional central limit theorem to capture the asymptotic behavior of the whole LDP-SGD path, thereby providing an asymptotically pivotal statistic for valid private inference.} 
   These results lay a solid foundation for real-time private decision-making, supported by theoretical guarantees.
\item  We conduct numerical experiments with simulated data to evaluate the performance of the proposed procedure in various settings, including varying cumulative sample size, different privacy budgets, and various models. Our findings indicate that the proposed procedure is computationally efficient, provides robust privacy guarantees, and significantly reduces data storage costs, all while maintaining high accuracy in estimation and inference. In addition, we apply the proposed method to the {the ride-sharing data and the US insurance data to demonstrate how the method can be useful in practice.}
\end{enumerate}

{The rest of this paper is organized as follows. In Section \ref{sec:2.1}, we introduce the key properties of DP. Section \ref{sec:2.2} outlines the problem formulation. In Section \ref{sec3}, we propose our LDP-SGD algorithm and provide its convergence rates. In Section \ref{sec3.2}, we establish a functional central limit theorem and introduce two online private inference methods: the private plug-in estimator and the random scaling method to construct asymptotically valid private confidence intervals. We evaluate the performance of the proposed procedure through simulation studies in Section \ref{sec4}. In Section \ref{sec5}, we apply the proposed method on two real-world datasets: the Ride-Sharing Data and the US Insurance Data. Ride-sharing data provides fare-related information over time, enabling analysis of market pricing dynamics in the ride-sharing industry. The US Insurance Data includes detailed personal and family information for one million clients, along with their associated insurance fees, offering valuable insights into insurance fee prediction.} Discussions are provided in Section \ref{sec:conclusion}. Some basic concepts of DP, the detailed description of private batch-means estimator, additional numerical results, and technical details are included in the Supplementary Material.


\section{Preliminaries} \label{sec:model}
\renewcommand{\theequation}{2.\arabic{equation}}
\setcounter{equation}{0}

In this section, we provide some preliminaries. To facilitate understanding, we first introduce the notations used throughout this paper. Let $\stackrel{D}{\rightarrow}$ denote convergence in distribution. For a $p$-vector $\boldsymbol{a} = (a_1,\ldots,a_{p})^\top\in\mathbb{R}^p$,   $\|\ve{a}\|_r = (\sum_{j=1}^{p}|a_j|^r)^{1/r}$ for $r\geq 1$. Define $\Lambda_1(\boldsymbol{A}),\ldots,\Lambda_p(\boldsymbol{A})$ as  the eigenvalues of a matrix $\boldsymbol{A}\in\mathbb{R}^{p\times p}$. Specifically, the largest and smallest eigenvalues are denoted by $\Lambda_{\max}(\boldsymbol{A})$ and $\Lambda_{\min}(\boldsymbol{A})$, respectively. We use $\|\boldsymbol{A}\|$ to denote the matrix operator norm of $\boldsymbol{A}$ and $\|\boldsymbol{A}\|_\infty$ to represent the element-wise $\ell_\infty$-norm of $\boldsymbol{A}$. For a square positive semi-definite matrix $\boldsymbol{D}$, we denote its trace by ${\rm tr}(\boldsymbol{D})$. For any two symmetric matrices $\boldsymbol{A}$ and $\boldsymbol{B}$ of the same dimension, $\boldsymbol{A} \preceq \boldsymbol{B}$ means that $\boldsymbol{B} - \boldsymbol{A}$ is a positive semi-definite matrix, while $\boldsymbol{A} \succeq \boldsymbol{B}$ means that $\boldsymbol{A} - \boldsymbol{B}$ is a positive semi-definite matrix. Let $\{a_n\}$ and $\{b_n\}$ be two sequences of positive numbers. Denote $a_n \lesssim b_n$ if there exists a positive constant $c_0$ such that $a_j/b_j\leq c_0$ for all $j \in \mathbb{N}^+$,  and $a_n \asymp b_n$ if there exist positive constants $c_1$, $c_2$ such that $c_1\leq a_j/b_j\leq c_2$ for all $j \in \mathbb{N}^+$.

\subsection{Some basic properties of differential privacy}\label{sec:2.1}

In this subsection, we only focus on some useful properties of DP, LDP, and Gaussian differential privacy (GDP), while more details about GDP-related concepts are provided in Section S1 of the Supplementary Material. Notice that a variety of basic algorithms can be made 
private by simply adding a properly scaled noise in the output. The scale of noise is characterized by the sensitivity of the algorithm. The formal definition is presented as follows:


{
\begin{definition}[Global sensitivity]
   For any (non-private) statistics $h(\boldsymbol{X})\in\mathbb{R}^p$ of the data set $\boldsymbol{X}$, the global sensitivity of $h$ is the (possibly infinite) number
   \begin{align*}
       {\rm{sens}}(h) = \sup\limits_{\boldsymbol{X},\boldsymbol{X}^\prime}\|h(\boldsymbol{X}) - h(\boldsymbol{X}^\prime)\|_2,
   \end{align*}
   where the supremum is taken over all pairs of data sets $\boldsymbol{X}$ and $\boldsymbol{X}^\prime$ that differ by one entry or datum.
\end{definition}
}

We then introduce the following mechanism to construct GDP estimators. Although only the univariate case was stated in Theorem 1 of \cite{dong2022gaussian}, its extension to general case can be obtained in the following proposition.

\begin{proposition}[Gaussian mechanism, \cite{dong2022gaussian}]\label{lemma1}
Define the Gaussian mechanism that operates on a statistic $h$ as $\tilde{h}(\boldsymbol{X}) = h(\boldsymbol{X}) + {\rm{sens}}(h)\boldsymbol{Z}/\mu$, where $\boldsymbol{Z}$ is a standard normal $p$-dimensional random vector and $\mu> 0$. Then, $\tilde{h}$ is $\mu$-GDP.
\end{proposition}

The post-processing and parallel composition properties are key aspects of DP, enabling the design of complex DP algorithms by combining simpler ones. Such properties are crucial in the algorithmic designs presented in later sections.

\begin{proposition}[Post-processing property, \cite{dong2022gaussian}]\label{post-p} Let $\mathcal{A}$ be an $\mu$-GDP algorithm, and $h$ be an arbitrary randomized algorithm which takes $\mathcal{A}(\boldsymbol{X})$ as input, then $h(\mathcal{A}(\boldsymbol{X}))$ is also $\mu$-GDP.
    
\end{proposition}

{
\begin{proposition}[Parallel composition, \cite{smith2021making}]\label{lemma2}
Let a sequence of $k$ mechanisms $\mathcal{A}_l$ each be $\mu_l$-GDP, $l=1,\ldots,k$. Let $\{\boldsymbol{X}_l\}_{l=1}^k$ be $k$ disjoint subsets of data. Define $\mathcal{O}_1=\mathcal{A}_1(\boldsymbol{X}_1)$ as the output of the mechanism $\mathcal{A}_1$, and the sequential output of the mechanism $\mathcal{A}_j$ is defined as $\mathcal{O}_j=\mathcal{A}_j(\boldsymbol{X}_j,\mathcal{O}_{j-1})$ for $j=2,\ldots,k$. Then the output of the last mechanism $\mathcal{O}_k$ is $\max\{\mu_1,\mu_2,\ldots,\mu_k\}$-GDP.
\end{proposition}
}

Since we will use a noisy version of the matrix to construct an asymptotic covariance matrix in a later section, we introduce the Matrix Gaussian mechanism. An intuitive approach to generating a differentially private matrix is to add independent and identically distributed (i.i.d.) noise to each individual component, as described below.

\begin{proposition}[Matrix Gaussian mechanism, \cite{avella2021differentially}]{\label{matrixgdp}} Let $\boldsymbol{G}\in\mathbb{R}^{n\times m}$ be a data matrix such that each row vector $\boldsymbol{g}_i$ satisfies $\|\boldsymbol{g}_i\|_2\leq 1$. Define the matrix Gaussian mechanism that operates on a statistic $h$ as $\tilde{h}(\boldsymbol{G}) = h(\boldsymbol{G}) + \boldsymbol{W}$, where $h(\boldsymbol{G}) = \boldsymbol{G}^\top \boldsymbol{G}/n$ and $\boldsymbol{W}$ is a symmetric random matrix whose upper-triangular elements, including the diagonal, are i.i.d. draws from $(2/(\mu n))\mathcal{N}(0,1)$. Then, $\tilde{h}$ is $\mu$-GDP.
\end{proposition}


\subsection{Problem formulation}\label{sec:2.2}

Consider $n\geq 2$, where $n$ observations, denoted by $\{\boldsymbol{z}_i\}_{i=1}^n$ with $\ve{z}_i = (\boldsymbol{x}_i^\top, y_i)^\top$ arrive sequentially. We assume these $n$ pairs of observations as i.i.d. copies of $(\boldsymbol{X}, Y)$ with $\boldsymbol{X}\in\mathbb{R}^p$ and $Y\in\mathbb{R}$ from a common cumulative distribution $F$. Our goal is to estimate a population parameter of interest $\boldsymbol{\theta}_0 \in \Theta \subseteq \mathbb{R}^p$, which can be formulated as the solution to the following optimization problem:
\begin{align}{\label{pop}}
    \boldsymbol{\theta}_0 =\operatorname{argmin}\left( L(\boldsymbol{\theta}) := {E}_{\mathcal{P}_{\ve{z}}}[\rho(\boldsymbol\theta, \boldsymbol{z})]=\int \rho(\boldsymbol\theta , \boldsymbol{z}) \mathrm{d} \mathcal{P}_{\ve{z}}\right),
\end{align}
where $\rho(\boldsymbol\theta,\ve{z})$ is some convex loss function with respect to $\boldsymbol{\theta}$, $\ve{z}$ is a random variable from the distribution $P_{\ve{z}}$, and $L(\boldsymbol\theta)$ is the population loss
function. 

In the offline setting, where the entire dataset is readily accessible, traditional deterministic optimization methods are commonly used to estimate the parameter $\boldsymbol{\theta}_0$ under the empirical distribution induced by the observed data.
However, for applications involving online data, where each sample arrives sequentially, storing all the data can be costly and inefficient. To address these challenges, the SGD algorithm \citep{robbins1951stochastic} is often used, especially for online learning, as it requires only one pass over the data. Starting from an initial point $\tilde{\boldsymbol\theta}_0$, the SGD algorithm recursively updates the estimate upon the arrival of each data point $\boldsymbol{z}_n$, $n\geq 1$, with the following update rule:
\begin{align}{\label{sgd}}
    \tilde{\boldsymbol\theta}_n =\tilde{\boldsymbol\theta}_{n-1} -\gamma_n\Psi(\tilde{\boldsymbol\theta}_{n-1},\boldsymbol{z}_n),  
\end{align}
where $\Psi({\boldsymbol\theta},\boldsymbol{z})$ denotes the stochastic gradient of $\rho({\boldsymbol\theta},\boldsymbol{z})$ with respect to the first argument ${\boldsymbol\theta}$, i.e., $\Psi({\boldsymbol\theta},\boldsymbol{z}) =\nabla \rho({\boldsymbol\theta},\boldsymbol{z})$ and $\gamma_n$ is the step size at the $n$th step. As suggested by \cite{ruppert1988efficient} and \cite{polyak1992acceleration}, we note that under some conditions, the averaging estimate
    $\tilde{{\boldsymbol\theta}}^\ast_n = \sum_{i=1}^n \tilde{{\boldsymbol\theta}}_i/n$
has the asymptotic normality
\begin{align*}
    \sqrt{n}(\tilde{{\boldsymbol\theta}}^\ast_n - {\boldsymbol\theta}_0)\stackrel{D}{\rightarrow} \mathcal{N}(\boldsymbol{0}, \boldsymbol\Sigma),
\end{align*}
where $\boldsymbol\Sigma = \ve{A}^{-1}\ve{S}\ve{A}^{-1}$, $\ve{A} = \nabla^2 L({\boldsymbol\theta}_0)$ is the Hessian matrix of $L({\boldsymbol\theta})$ at ${\boldsymbol\theta} = {\boldsymbol\theta}_0$, and $\ve{S}$ is the covariance matrix of $\nabla \rho({\boldsymbol\theta}_0,\ve{z})$, given by $\ve{S}= {E}\{\nabla \rho({\boldsymbol\theta}_0,\ve{z})\nabla \rho({\boldsymbol\theta}_0,\ve{z})^\top\}$.

A significant limitation of traditional SGD is its vulnerability to privacy breaches, arising from the extensive data and gradient queries made at each iteration. In this paper, our work focuses primarily on settings where trusted data collectors are not required under the constraints of LDP. In such settings, each individual is required to locally apply a DP mechanism to their data before transmitting the perturbed data to the server. 
In particular, we introduce an LDP-SGD algorithm designed to fit the model in (\ref{pop}) and accommodate timely online private statistical inference for $\boldsymbol{\theta}_0$, thereby providing rigorous individual-level data privacy protection while reducing both time and space complexity.

\section{Methodology}\label{sec3}


In this subsection, we introduce an LDP estimator using noisy SGD algorithms within the framework of streaming data. The heightened risk of privacy leakage arises from the numerous data and gradient queries inherent in the SGD algorithm. To facilitate our work, we impose a restriction on the class of estimators and enforce the following uniform boundedness condition.
\begin{condition}{\label{Con1}}
The gradient of the loss function $\rho$ satisfies 
\begin{align*}
    \sup\limits_{\boldsymbol\theta\in\Theta, \ve{z}\in\mathcal{Z}}\|\Psi(\boldsymbol\theta, \ve{z})\|_2\leq B_0
\end{align*}
for some positive constant $B_0$.
 \end{condition}

\begin{figure}[h]

  \centering
 \includegraphics[width=4.0cm,height=5.0cm]{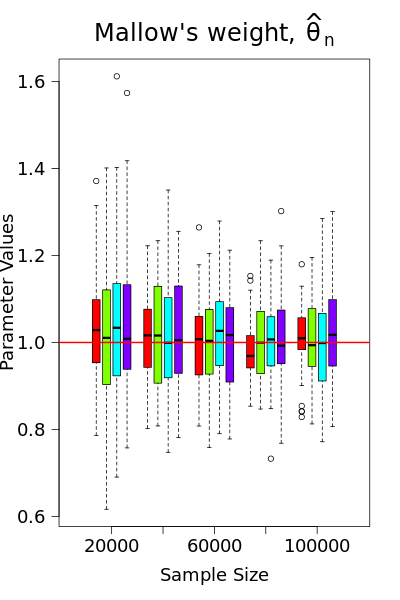}\includegraphics[width=4.0cm,height=5.0cm]{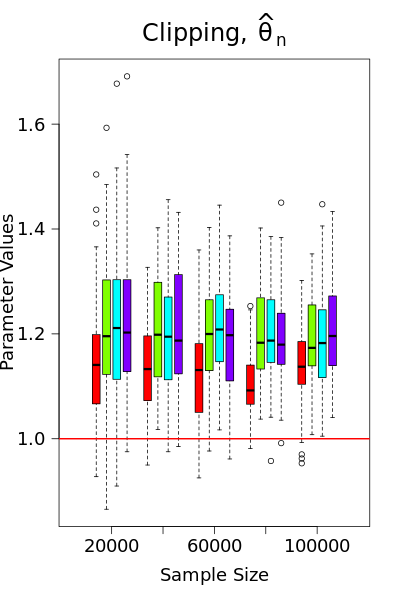}
\includegraphics[width=4.0cm,height=5.0cm]{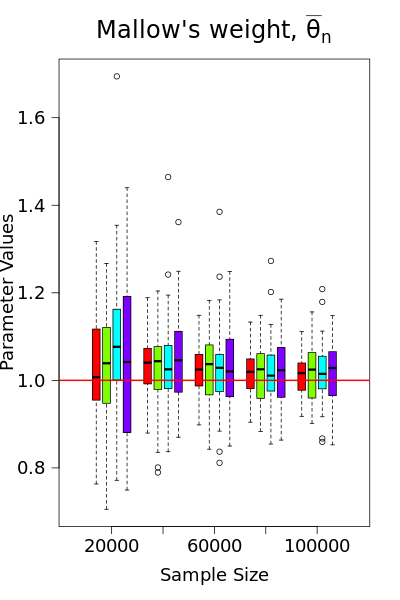}\includegraphics[width=4.0cm,height=5.0cm]{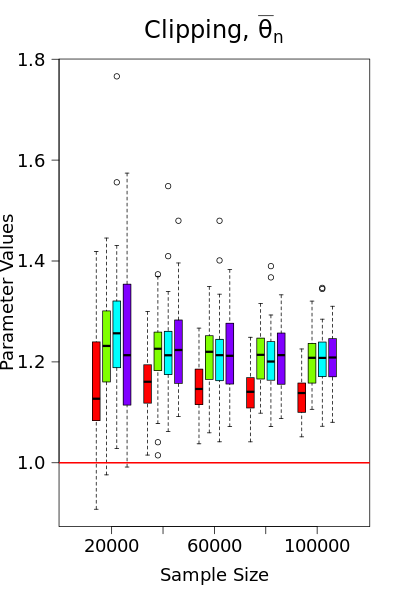}
  \caption{Boxplots comparing estimators from Mallow's weights and gradient clipping for logistic regression with dimension $p=3$. The red line indicates the true parameter value, $\theta^*=1$, while the boxplots represent the distribution of estimators across 50 replications. The four colors correspond to four different parameters. Clipping results in a positive bias for logistic regression, which does not arise using Mallow's weights.} 
  \label{fig:clip}
\end{figure}

Condition \ref{Con1} is satisfied by using either Mallow's weights (as discussed in Examples 1-3) or gradient clipping. However, as demonstrated in \cite{song2021evading, avella2021differentially}, gradient clipping is not ideal from a statistical standpoint, as it may result in inconsistent estimators. In Figure \ref{fig:clip}, we compare the performance of gradient clipping and Mallow's weights using simulated logistic regression data. The results show that the parameter estimates from gradient clipping exhibit bias that does not shrink toward zero with the sample size, further underscoring our preference for Mallow's weights, alongside the assumption of a bounded gradient. 
Meanwhile, Condition \ref{Con1} implies that the loss function $\rho$ has a known bound on the gradient $\Psi(\boldsymbol\theta, \ve{z})$, which ensures that $2B_0$ is an upper bound for the global sensitivity of $\Psi(\boldsymbol\theta, \ve{z})$. At the $n$th iteration with observed data point $\ve{z}_n$, $n\geq 1$, we consider the following locally private SGD estimator with the noisy version of the iterates (\ref{sgd}){
\begin{align}{\label{psgd}}
    \hat{\boldsymbol\theta}_n =\hat{\boldsymbol\theta}_{n-1} -\gamma_n\left\{\Psi(\hat{\boldsymbol\theta}_{n-1},\ve{z}_n) + \frac{2 B_0}{\mu_n}\boldsymbol\xi_n\right\}, 
\end{align}
where the final estimate is the averaging estimate denoted by $\bar{{\boldsymbol\theta}}_n = \sum_{i=1}^n\hat{\boldsymbol\theta}_i/n$ and $\{\boldsymbol\xi_n\}_{n\geq 1}$ is a sequence of i.i.d. standard $p$-dimensional Gaussian random vectors. Unlike the standard SGD algorithm, the locally private SGD algorithm introduces calibrated noise to the gradient computations at each step of the model parameter update process, ensuring rigorous privacy protection. In particular, taking $B_0=0$ in the iterates (\ref{psgd}) recovers the standard SGD algorithm in (\ref{sgd}). We call the proposed iterates in (\ref{psgd}) as the LDP-SGD algorithm. The following proposition shows that our proposed LDP-SGD algorithm is $\max\{\mu_1,\ldots,\mu_n\}$-GDP. 

\begin{proposition}{\label{alg-ave}}
Given an initial estimate $\hat{\boldsymbol{\theta}}_{0}\in\mathbb{R}^p$, consider the iterates $\{\hat{\boldsymbol\theta}_n\}_{n\geq 1}$ defined in (\ref{psgd}). Then the final output $\bar{\boldsymbol{\theta}}_n$ is $\max\{\mu_1,\ldots,\mu_n\}$-GDP.
\end{proposition}


From (\ref{psgd}), each individual has different privacy budgets $\mu_n$, $n\geq 1.$ 
A special case arises when the proposed estimator reduces to  $\mu$-GDP with $\mu = \mu_1 = \cdots = \mu_n$, thereby not considering different privacy budgets for each individual. The techniques needed to handle the cases of varing $\mu_n$ are the same as for the case of the common $\mu$. For simplicity, we assume that  $\mu=\mu_1=\cdots=\mu_n$ for each individual. 

\begin{condition}{\label{Con2}}
Assume that the objective function $L({\boldsymbol\theta})$ is differentiable, $\zeta$-smooth, and $\lambda$-strongly convex, in the sense
\begin{align*}
    & L({\boldsymbol\theta}_1) - L({\boldsymbol\theta}_2) \leq \langle \nabla L({\boldsymbol\theta}_2), {\boldsymbol\theta}_1-{\boldsymbol\theta}_2 \rangle + \frac{\zeta}{2}\|{\boldsymbol\theta}_1-{\boldsymbol\theta}_2\|^2,\quad \forall{\boldsymbol\theta}_1, {\boldsymbol\theta}_2\in{\Theta}\subseteq\mathbb{R}^p,\\
    & L({\boldsymbol\theta}_1) - L({\boldsymbol\theta}_2) \geq \langle \nabla L({\boldsymbol\theta}_2), {\boldsymbol\theta}_1-{\boldsymbol\theta}_2 \rangle + \frac{\lambda}{2}\|{\boldsymbol\theta}_1-{\boldsymbol\theta}_2\|^2,\quad \forall{\boldsymbol\theta}_1, {\boldsymbol\theta}_2\in{\Theta}\subseteq\mathbb{R}^p.
\end{align*}
 \end{condition}
Strong convexity and smoothness are standard conditions for the convergence analysis of (stochastic) gradient optimization methods. Similar conditions can be found in \cite{ vaswani2022towards,zhu2023online}. Due to space limitations, several useful properties of strongly convex and smooth functions are provided in Lemma S1 of the Supplementary Material. As stated in Lemma S1, the strong convexity and smoothness of $L({\boldsymbol\theta})$ imply that $\Lambda_{\min}\{\nabla^2 L({\boldsymbol\theta})\}\geq \lambda$ and $\Lambda_{\max}\{\nabla^2 L({\boldsymbol\theta})\}\leq \zeta$, respectively.  

\begin{condition}{\label{Con3}}
Let $\ve{\eta}_i = \nabla L(\hat{\boldsymbol\theta}_{i-1}) - \Psi(\hat{\boldsymbol\theta}_{i-1}, \ve{z}_{i})$ denote the gradient noise. There exists some positive constant $C_1$ such that the conditional covariance of $\ve{\eta}_n$ has an expansion around $\ve{S}$, satisfies 
\begin{align*}
   & \|E(\ve{\eta}_n\ve{\eta}^\top_n |\mathcal{F}_{n-1}) - \ve{S}\| \leq C_1(\|\hat{\boldsymbol{\delta}}_{n-1}\|_2 + \|\hat{\boldsymbol{\delta}}_{n-1}\|_2^2),\\& |{\rm tr}\{E(\ve{\eta}_n\ve{\eta}^\top_n |\mathcal{F}_{n-1}) - \ve{S}\}| \leq C_1(\|\hat{\boldsymbol{\delta}}_{n-1}\|_2 + \|\hat{\boldsymbol{\delta}}_{n-1}\|_2^2),
\end{align*}
where $\mathcal{F}_{n-1}$ is the $\sigma$-algebra generated by $\{\ve{z}_1,\ldots,\ve{z}_{n-1}\}$ and $\hat{\ve\delta}_n = \hat{\ve\theta}_n - {\boldsymbol\theta}_0$ is the error sequence. In addition, the fourth conditional moment of $\ve{\eta}_n$ is bounded by 
\begin{align*}
    E(\|\ve{\eta}_n\|_2^4 | \mathcal{F}_{n-1})\leq C_2(1 + \|\hat{\boldsymbol\delta}_{n-1}\|_2^4)
\end{align*}
for some positive constant $C_2.$
\end{condition}

Condition \ref{Con3} is a mild requirement on the loss function. As shown in Lemma 3.1 of \cite{chen2020statistical}, Condition \ref{Con3} holds if the Hessian matrix of $\rho({\boldsymbol\theta},\ve{z})$ is bounded by some random function $H(\ve{z})$ with a bounded fourth moment. Examples that satisfy this condition include the robust estimation methods for commonly used linear and logistic regressions. With the aforementioned assumptions, we now present the following error bounds on the private SGD iterates. 

\begin{theorem}{\label{order}}
  Under Conditions \ref{Con1}-\ref{Con3}, there exist some positive constants $n_0\in\mathbb{N}$ and $c_p$ that depends on the dimension $p$, such that for $n\geq n_0$,   $\hat{\boldsymbol\delta}_n = \hat{\boldsymbol\theta}_n - {\ve\theta}_0$ satisfies the following 
  \begin{align*}
      E(\|\hat{\boldsymbol\delta}_n\|_2^m)\lesssim n^{-m\alpha/2}\{(\gamma c_pB_0^2/\mu^2)^{m/2}/\lambda + \|\hat{\boldsymbol\delta}_0\|_2^m\},\quad m=1,2,4,
  \end{align*}
  when the step-size is chosen to be $\gamma_i =\gamma i^{-\alpha}$ with $\gamma>0$ and $1/2<\alpha<1$.
\end{theorem}

Theorem \ref{order} provides simplified bounds on the conditional moments of $\hat\delta_n$,  including the first, second, and fourth moments. These results characterize the convergence rate of the last-step estimator $\hat{\boldsymbol{\theta}}_n$. 
To illustrate the applicability of our proposed algorithm, we provide several examples below. Our framework, as it turns out, accommodates a wide range of important statistical models.

\begin{example}[Linear Regression]\label{exam1}
   Let $\boldsymbol{z}_n = (\boldsymbol{x}^\top_n,y_n)^\top$, $n=1,2,\ldots,$ be a sequence of i.i.d. copies from $(\ve{X}^\top,Y)^\top$ with $Y\in\mathbb{R}$ and $\ve{X}\in\mathbb{R}^p$, satisfying $y_n = \boldsymbol{x}_n^\top\boldsymbol{\theta}_0 + \varepsilon_n$, where $\varepsilon_n\in\mathbb{R}$ is the random noise. The loss function can be chosen as $\rho(\boldsymbol{\theta},\boldsymbol{z})= h_c(y - \ve{x}^\top{\boldsymbol\theta})\omega(\ve{x})$, 
where $h_c(\cdot)$ is the Huber loss with truncated parameter $c$ and $\omega(\ve{x}) = \min(1, 2/\|\ve{x}\|_2^2)$ downweights outlying covariates. With this construction, it is straightforward to verify that the global sensitivity of $\Psi(\boldsymbol{\theta},\boldsymbol{z})$ is $2\sqrt{2}c$. Letting $B_0 = \sqrt{2}c$, our proposed LDP-SGD updates for ${\boldsymbol\theta}_0$, as defined in (\ref{psgd}) is given by:
\begin{align*}
    \hat{\boldsymbol\theta}_n =\hat{\boldsymbol\theta}_{n-1} + \gamma_n\{\min(1,{c}/{|y_n - \ve{x}_n^\top{\hat{\boldsymbol\theta}_{n-1}}|})\ve{x}_n(y_n -\ve{x}_n^\top{\hat{\boldsymbol\theta}_{n-1}})\omega(\ve{x}_n)\} + \frac{2\sqrt{2}c\gamma_n}{\mu_n}\boldsymbol\xi_n,~~ n\geq 1, 
\end{align*}
where $I(\cdot)$ is the indicator function. 
\end{example}

\begin{example}[Logistic Regression]\label{exam2}
   An important data science problem is logistic regression for binary classification problems. Specifically, let $ \boldsymbol{z}_n = (\ve{x}_n^\top,y_n)^\top$, $n=1,2,\ldots,$ be a sequence of i.i.d. copies from $(\ve{X}^\top,Y)^\top$ with $Y\in\{-1,1\}$ and $\ve{X}\in\mathbb{R}^p$, satisfying $y_n\sim {\rm Bernoulli}(\{1 + \exp(-\boldsymbol{x}_n^\top\boldsymbol{\theta}_0)\}^{-1})$. Taking Mallow's weighted version of the usual cross-entropy loss as $\rho(\boldsymbol{\theta},\boldsymbol{z}) = \{-y\log(1/\{1 + \exp(\boldsymbol{x}^\top\boldsymbol{\theta})\}) + (1-y)\log(\exp(\boldsymbol{x}^\top\boldsymbol{\theta})/\{1 + \exp(\boldsymbol{x}^\top\boldsymbol{\theta})\})\}\omega(\ve{x})$, where $\omega(\ve{x})$ is defined in the previous example.
  By this construction, we can easily verify that the global sensitivity of $\Psi(\boldsymbol{\theta},\boldsymbol{z})$ is $2\sqrt{2}$. Setting $B_0 = \sqrt{2}$, our proposed LDP-SGD updates for ${\boldsymbol\theta}_0$, as defined in (\ref{psgd}) is given by:
\begin{align*}
    \hat{\boldsymbol\theta}_n =\hat{\boldsymbol\theta}_{n-1} + \gamma_n\ve{x}_n(y_n -\{1+\exp(-\ve{x}_n^\top{\hat{\boldsymbol\theta}_{n-1}})\}^{-1})\omega(\ve{x}_n)+ \frac{2\sqrt{2}\gamma_n}{\mu_n}\boldsymbol\xi_n, ~~ n\geq 1.
\end{align*}
\end{example}

\begin{example}[Robust Expectile Regression]\label{exam3}
As an important extension of linear regression, robust expectile regression allows the regression coefficients $\boldsymbol\theta_0(\tau)$ to vary across different
values of location parameter $\tau$, and thereby offers insights into the entire conditional distribution of $Y$ given $\ve{X}$; see \cite{man2024retire}.  Specifically, given a location
parameter $\tau\in(0,1)$, let $ \boldsymbol{z}_n = (\ve{x}_n^\top,y_n)^\top$, $n=1,2,\ldots,$ be a sequence of i.i.d. copies from $(\ve{X}^\top,Y)^\top$, satisfying  $y_n = \boldsymbol{x}_n^\top\boldsymbol{\theta}_0(\tau) + \varepsilon_n(\tau)$. The loss function can be chosen as $\rho(\boldsymbol{\theta},\boldsymbol{z})= |\tau- I(y-\boldsymbol{x}^\top\boldsymbol{\theta}<0)|\cdot h_c(y-\boldsymbol{x}^\top\boldsymbol{\theta})\omega(\ve{x})$, where $h_c(\cdot)$ is the Huber loss with truncated parameter $c$. With this construction, it is straightforward to verify that the global sensitivity of $\Psi(\boldsymbol{\theta},\boldsymbol{z})$ is $2\sqrt{2}c\max(\tau,1-\tau)$. Taking $B_0 = \sqrt{2}c\max(\tau,1-\tau)$, our proposed LDP-SGD updates for ${\boldsymbol\theta}_0$, as defined in (\ref{psgd}) is given by:
\begin{align*}
    \hat{\boldsymbol\theta}_n & =\hat{\boldsymbol\theta}_{n-1} + \gamma_n|\tau- I(y_n-\boldsymbol{x}^\top_n\hat{\boldsymbol\theta}_{n-1}<0)|\cdot \min(1,{c}/{|y_n - \ve{x}_n^\top{\hat{\boldsymbol\theta}_{n-1}}|}) \ve{x}_n(y_n -\ve{x}_n^\top{\hat{\boldsymbol\theta}_{n-1}})\omega(\ve{x}_n) \\
    & \quad + \frac{2\sqrt{2}c\max(\tau,1-\tau)\gamma_n }{\mu_n}\boldsymbol\xi_n,~~ n\geq 1. 
\end{align*}
\end{example}

\section{Online private inference}\label{sec3.2}

{In this subsection, we introduce two online inference methods, distinguished by the availability of second-order information from the loss function. One is the online private plug-in approach that directly uses the Hessian information and the other is the random scaling that utilizes only the information from the path of our proposed LDP-SGD iteration in (\ref{psgd}). In addition to these two methods, we also consider a private batch-means estimator that utilizes the iterations generated by the LDP-SGD algorithm to estimate the asymptotic covariance. The detailed construction of this estimator, along with a comparative discussion of its theoretical properties relative to the above two methods, is provided in the Supplementary Material.

To elucidate these methods, we first examine the asymptotic behavior of the entire LDP-SGD path rather than its simple average $\bar{\boldsymbol\theta}_n$. Specifically, we present the functional central limit theorem, which shows that the standardized partial-sum process converges in distribution to a rescaled Brownian motion.
}


\begin{theorem}[Functional CLT]{\label{fclt}}
   Consider LDP-SGD iterates in (\ref{psgd}) with step-size $\gamma_i=\gamma i^{-\alpha}$ for some constant $\gamma >0$ and $1/2<\alpha<1$. Then, under Conditions \ref{Con1}-\ref{Con3}, the following random function weakly converges to a scaled Brownian motion, i.e.,
    \begin{align*}
  \phi_n(r):=  \frac{1}{\sqrt{n}}\sum\limits_{i=1}^{\lfloor nr \rfloor}(\hat{\boldsymbol\theta}_i - \boldsymbol{\theta}_0) \Rightarrow \boldsymbol{A}^{-1}\{\boldsymbol{S} + (4B^2_0/\mu^2)\boldsymbol{I}\}^{1/2}\mathcal{W}_p(r)~~{\rm{for}}~~r\in(0,1]
\end{align*}
as $n\rightarrow \infty$, where $\ve{A} = \nabla^2 L({\boldsymbol\theta}_0)$,  $\ve{S}={E}\{\nabla \rho({\boldsymbol\theta}_0,\ve{z})\nabla \rho({\boldsymbol\theta}_0,\ve{z})^\top\}$, $\mathcal{W}_p$ is the $p$-dimensional vector of the independent standard Wiener processes on $[0,1]$, and the symbol $\Rightarrow$ stands for the weak convergence.
\end{theorem}

The result derived from this theorem is stronger than the asymptotic normality of the proposed private estimator $\bar{\boldsymbol\theta}_n$ under the stated assumptions. In particular, by applying the continuous mapping theorem to Theorem \ref{fclt}, we are able to directly establish the following corollary.

\begin{corollary}{\label{the1}}
    Under the conditions  of Theorem \ref{fclt}, the averaging estimator $\bar{\boldsymbol\theta}_n = \sum_{i=1}^n\hat{\boldsymbol\theta}_i/n$ satisfies 
$\sqrt{n}(\bar{\boldsymbol\theta}_n - {\boldsymbol\theta}_0)\stackrel{D}{\rightarrow} \mathcal{N}(\boldsymbol{0}, \widetilde{\boldsymbol\Sigma})$,~ where $ \widetilde{\boldsymbol\Sigma} = {\ve{A}}^{-1}\tilde{\boldsymbol{S}}\ve{A}^{-1}$, and  $\tilde{\boldsymbol{S}} = \ve{S} + (4B_0^2/\mu^2)\ve{I}$.
\end{corollary}
In contrast to the covariance matrix of the non-private estimator in (\ref{sgd}), the additional term $(4B_0^2/\mu^2){\ve{I}}$ in the covariance matrix of Corollary \ref{the1} represents the ``cost of privacy'' for our proposed LDP-SGD algorithm.

\subsection{Private plug-in estimator}
In the previous subsection, we present the asymptotic distribution of the proposed LDP-SGD estimator in Corollary \ref{the1}. For the purpose of conducting statistical inference of $\boldsymbol{\theta}_0$, a consist estimator of the limiting covariance matrix $\widetilde{\boldsymbol\Sigma}$ from Corollary \ref{the1} is required. An intuitive way to constructing confidence intervals for $\boldsymbol{\theta}_0$ is to estimate each component of the sandwich formula $\widetilde{\boldsymbol\Sigma}$ separately. Without privacy constraints, the estimators for $\ve{A}$ and $\tilde{\boldsymbol{S}}$ are given by:
\begin{align*}
    \ve{A}_n = \frac{1}{n}\sum\limits_{i=1}^n \dot{\Psi}(\hat{\boldsymbol\theta}_{i-1}, \ve{z}_i)\quad{\rm{and}}\quad \tilde{\boldsymbol{S}}_n = \frac{1}{n}\sum\limits_{i=1}^n\Psi(\hat{\boldsymbol\theta}_{i-1},\ve{z}_i)\Psi(\hat{\boldsymbol\theta}_{i-1},\ve{z}_i)^\top + \frac{4B_0^2}{\mu^2}\ve{I}
\end{align*}
as long as the information of $\dot{\Psi}(\hat{\boldsymbol\theta}_{i-1},\ve{z}_i)$ is available, where $\dot{\Psi}(\boldsymbol{\theta},\boldsymbol{z})$ denotes the derivative of $\Psi(\boldsymbol{\theta},\boldsymbol{z})$ with respect to its first argument $\boldsymbol{\theta}$. Notice that each summand in both $\ve{A}_n$ and $\tilde{\boldsymbol{S}}_n$ involves different $\hat{\boldsymbol\theta}_i$. This allows computations to be carried out in an online manner without the need to store the entire dataset. The consistency of the corresponding non-private plug-in estimator is also established and provided in the Lemma S8 of the Supplementary Material.

However, in the DP setting, direct application of this plug-in construction is not feasible, as neither ${\boldsymbol{A}}_n$ nor $\tilde{\boldsymbol{S}}_n$ is differentially private. To overcome this limitation, we propose constructing DP versions of these matrices using the Matrix Gaussian mechanism, as outlined in Proposition \ref{matrixgdp}. This requires the Hessian to have a specific factorization structure that facilitates the effective application of Proposition \ref{matrixgdp}. Moreover, we assume that the spectral norm of the Hessian is uniformly bounded, as stated below.
\begin{condition}{\label{Con5}}
    The Hessian $\nabla^2{\rho}(\boldsymbol\theta,\boldsymbol{z})$ is positive definite for all $\boldsymbol\theta\in\mathbb{R}^p$ with the form $\nabla^2{\rho}(\boldsymbol\theta,\boldsymbol{z}) = m(\boldsymbol\theta,\boldsymbol{z})m(\boldsymbol\theta,\boldsymbol{z})^\top$, where $m:\Theta \times \mathcal{Z}\rightarrow \mathbb{R}^p$ and $\sup_{\boldsymbol\theta,\boldsymbol{z}}\|m(\boldsymbol\theta,\boldsymbol{z})\|_2^2 \leq B_1 <\infty$.
\end{condition}
We are now ready to present our DP counterparts of $\boldsymbol{A}_n$ and $\tilde{\boldsymbol{S}}_n$, which appear in the plug-in estimator. Consider the following private estimators:
\begin{align*}
    &\hat{\boldsymbol{A}}_n =\frac{1}{n}\sum\limits_{i=1}^n \dot{\Psi}(\hat{\boldsymbol\theta}_{i-1}, \boldsymbol{z}_i)  + \frac{2B_1}{n\mu}\boldsymbol{M}_{1i} = \boldsymbol{A}_n + \frac{2B_1}{n\mu}\boldsymbol{M}_{1},\\
    &\hat{\boldsymbol{S}}_n = \frac{1}{n}\sum\limits_{i=1}^n\Psi(\hat{\boldsymbol\theta}_{i-1},\boldsymbol{z}_i)\Psi(\hat{\boldsymbol\theta}_{i-1},\boldsymbol{z}_i)^\top + \frac{4B_0^2}{\mu^2}\boldsymbol{I} + \frac{2B_0^2}{n\mu}\boldsymbol{M}_{2i} = \tilde{\boldsymbol{S}}_n +\frac{2B_0^2}{n\mu}\boldsymbol{M}_{2} ,
\end{align*}
where $\boldsymbol{M}_{1}$ and $\boldsymbol{M}_{2}$ are i.i.d. symmetric random matrices whose upper-triangular elements, including the diagonals, are i.i.d. standard normal. However, the DP matrices $\hat{\boldsymbol{A}}_n$ and $\hat{\boldsymbol{S}}_n$ can potentially be negative definite. To address this concern, we apply a truncation procedure to the eigenvalues of the matrices. Specifically, we first perform spectral decompositions: $\hat{\boldsymbol{A}}_n = \boldsymbol{\Gamma}_A \boldsymbol{D}_A \boldsymbol{\Gamma}_A^{\top}$ and $\hat{\boldsymbol{S}}_n = \boldsymbol{\Gamma}_S \boldsymbol{D}_S \boldsymbol{\Gamma}_S^{\top}$, where $\boldsymbol{\Gamma}_A$ and $\boldsymbol{\Gamma}_S$ are orthogonal matrices. Given fixed threshold parameters $\kappa_1$ and $\kappa_2$, we define the truncated estimators as $\hat{\boldsymbol{A}}^{\ast}_n = \boldsymbol{\Gamma}_A \hat{\boldsymbol{D}}_A \boldsymbol{\Gamma}_A^{\top}$ and $\hat{\boldsymbol{S}}^{\ast}_n = \boldsymbol{\Gamma}_S \hat{\boldsymbol{D}}_S \boldsymbol{\Gamma}_S^{\top}$, where the diagonal elements are thresholded as $(\hat{\boldsymbol{D}}_A)_{i,i} = \max\{\kappa_1, (\boldsymbol{D}_A)_{i,i}\}$ and $(\hat{\boldsymbol{D}}_S)_{i,i} = \max\{\kappa_2, (\boldsymbol{D}_S)_{i,i}\}$ for $i = 1, \ldots, p$. In other words, eigenvalues smaller than $\kappa_1$ or $\kappa_2$ are truncated to ensure numerical stability and positive semi-definiteness. The DP sandwich estimator can then be constructed as follows:
\begin{align}{\label{plug-in}}
    \widehat{\boldsymbol\Sigma}_n = \hat{\boldsymbol{A}}^{\ast-1}_n\hat{\boldsymbol{S}}^\ast_n \hat{\boldsymbol{A}}^{\ast-1}_n.
\end{align}
To establish
consistency of our proposed private plug-in estimator $\widehat{\boldsymbol\Sigma}_n$, we need the following additional smoothness assumption regarding the Hessian, similar to those found in the
literature \citep{chen2020statistical,chen2024online}. 
\begin{condition}{\label{hess-lip}}
Assume that for all $\boldsymbol\theta\in\mathbb{R}^p$,  the following holds:
\begin{align*}
& E\|\nabla^2{\rho}({\boldsymbol\theta},\ve{z}) - \nabla^2{\rho}({\boldsymbol\theta}_0,\ve{z})\| \leq K \|{\boldsymbol\theta} - {\boldsymbol\theta}_0\|_2,\\
&\|E\{\nabla^2{\rho}({\boldsymbol\theta}_0,z)\}^2 - \ve{A}^2\|\leq C_3,
\end{align*}
where $K$ and $C_3$ are some postive constants. Furthermore, assume that $\Lambda_{\min}(\ve{A}) >\delta.$
\end{condition}
\begin{theorem}[Error rate of the private plug-in estimator]\label{plug1}
   Under the conditions of Corollary \ref{the1}, and Conditions \ref{Con5}-\ref{hess-lip}, then  $\widehat{\boldsymbol\Sigma}_n$ is $\sqrt{3}\mu$-GDP and converges to the asymptotic covariance matrix that satisfies the following 
   \begin{align*}
        E\| \widehat{\boldsymbol\Sigma}_n - \boldsymbol{A}^{-1}\tilde{\boldsymbol{S}}\boldsymbol{A}^{-1}\| \lesssim \|\boldsymbol{S}\|\frac{\sqrt{\gamma c_p}B^2_0 K}{\mu}\left(1 + \Lambda_{\min}(\boldsymbol{S}) +\frac{4B_0^2}{\mu^2}\right)\left(n^{-\alpha/2} + \frac{\sqrt{\gamma c_p}K}{\mu}n^{-\alpha}\right),
    \end{align*}
    when the step-size is chosen to be $\gamma_i = \gamma i^{-\alpha}$ with $\gamma>0$ and $\alpha\in(1/2,1).$ 
\end{theorem}
Theorem \ref{plug1} establishes the consistency of the proposed private plug-in estimator $\widehat{\boldsymbol\Sigma}_n$. Consequently, 
we can consider the following $\sqrt{3}\mu$-GDP $(1-\alpha_0)$-confidence interval for $\theta_{0,j}$, $j=1,\ldots,p$:
\begin{align*}
    {\rm{CI}}_{1-\alpha_0}^{\rm{plug}}(\theta_{0,j}) = \left[\bar{\theta}_{n,j} - \hat{\sigma}_{n,j}^P z_{1-\alpha_0/2}/\sqrt{n}, \bar{\theta}_{n,j} + \hat{\sigma}_{n,j}^P z_{1-\alpha_0/2}/\sqrt{n} \right],
\end{align*}
where $\hat{\sigma}_{n,j}^P = (\widehat{\boldsymbol\Sigma}_n)^{1/2}_{j,j}$ and $z_{1-\alpha_0/2}$ is the ($1-\alpha_0/2$)-quantile of the standard normal distribution. In particular, we have the following corollary, which demonstrates that ${\rm{CI}}_{1-\alpha_0}^{\rm{plug}}(\theta_{0,j})$ is an asymptotically exact confidence interval. 
\begin{corollary}{\label{plug-cor}}
    Under the conditions of Theorem \ref{plug1}, when $p$ is fixed and $n\rightarrow \infty$, we have 
    \begin{align*}
        {\rm{Pr}}\left(\bar{\theta}_{n,j} - z_{1-\alpha_0/2}\hat{\sigma}_{n,j}^P/\sqrt{n} \leq \theta_{0,j}\leq\bar{\theta}_{n,j} + z_{1-\alpha_0/2}\hat{\sigma}_{n,j}^P/\sqrt{n} \right) \rightarrow 1-\alpha_0.
    \end{align*}
\end{corollary}

\subsection{Random scaling}
Despite the simplicity of the private plug-in approach, the proposed estimator $\widehat{\boldsymbol{\Sigma}}_n$ incurs additional computational and storage cost,  primarily because it requires extra function-value queries to construct $\hat{\boldsymbol{A}}^\ast_n$. To mitigate this issue, we propose the random scaling inference procedure, which avoids the direct estimation of the asymptotic variance. Instead, the procedure studentizes $\bar{\boldsymbol\theta}_n$ using a matrix derived from iterations along the LDP-SGD path. Specifically, with Theorem \ref{fclt}, we observe that $\phi_n(1) = \sum_{i=1}^n(\hat{\boldsymbol\theta}_i -\boldsymbol{\theta}_0)/\sqrt{n} = \sqrt{n}(\bar{\boldsymbol\theta}_n - \boldsymbol\theta_0)$. Consequently, $\phi_n(r) - \lfloor nr \rfloor\phi_n(1)/{n} = \sum_{i=1}^{\lfloor nr \rfloor}(\hat{\boldsymbol\theta}_i - \bar{\boldsymbol\theta}_n)/\sqrt{n}$ eliminates the dependence on $\boldsymbol{\theta}_0$. To remove the dependence on the unknown scale  $\boldsymbol{A}^{-1}\{\boldsymbol{S} + (4B^2_0/\mu^2)\boldsymbol{I}\}^{1/2}$, we studentize $\phi_n(1)$ via
\begin{align*}
    \hat{\boldsymbol{V}}_n =&\frac{1}{n}\sum\limits_{b=1}^n\left\{\phi_n(b/n) - \frac{b}{n}\phi_n(1)\right\}\left\{\phi_n(b/n) - \frac{b}{n}\phi_n(1)\right\}^\top\\
    =&\frac{1}{n}\sum\limits_{b=1}^n\left\{\frac{1}{\sqrt{n}}\sum\limits_{i=1}^b (\hat{\boldsymbol\theta}_i - \bar{\boldsymbol\theta}_n)\right\}\left\{\frac{1}{\sqrt{n}}\sum\limits_{i=1}^b (\hat{\boldsymbol\theta}_i - \bar{\boldsymbol\theta}_n)\right\}^\top.  
\end{align*}
This results in the statistic $\phi^\top_n(1)\hat{\boldsymbol{V}}_n^{-1}\phi_n(1)$ being asymptotically pivotal, with a distribution that is free of any unknown nuisance parameters. This conclusion is supported by the following corollary, which directly follows from Theorem \ref{fclt} and the continuous mapping theorem. Thus, the proof of this corollary is omitted.

\begin{corollary}{\label{fclt_cor1}}
    Under the conditions of Theorem \ref{fclt}, we have
    \begin{align*}
      \phi^\top_n(1)\hat{\boldsymbol{V}}_n^{-1}\phi_n(1) \stackrel{D}{\rightarrow} \mathcal{W}_p(1)^\top \left\{\int_0^1 \bar{\mathcal{W}}(r)\bar{\mathcal{W}}(r)^\top {\rm d}r\right\}^{-1}\mathcal{W}_p(1),
    \end{align*}
    where $\bar{\mathcal{W}}(r) = \mathcal{W}_p(r) -r \mathcal{W}_p(1)$.
\end{corollary}
This corollary implies that $\phi^\top_n(1)\hat{\boldsymbol{V}}_n^{-1}\phi_n(1)$ can be used to construct valid asymptotic confidence intervals. From Corollary \ref{fclt_cor1}, we know that the $t$-statistic $\sqrt{n}(\bar{\theta}_{n,j} - \theta_{0,j})$, $j=1,\ldots,p$, is asymptotically pivotal and converges to the following pivotal limiting distribution  
\begin{align}{\label{limi-dis}}
    \frac{\sqrt{n}(\bar{\theta}_{n,j} - \theta_{0,j})}{\sigma^R_{n,j}} \stackrel{D}{\rightarrow} \frac{\mathcal{W}_1(1)}{[\int_0^1\{\mathcal{W}_1(r) - r\mathcal{W}_1(1)\}^2{\rm d}r ]^{1/2}},
\end{align}
where $\sigma^R_{n,j} = ({\hat{\boldsymbol{V}}_{n}})_{j,j}^{1/2}$. Notice that the random matrix $\hat{\boldsymbol{V}}_n$ can be updated in an online fashion, thereby enabling the construction of confidence intervals. Specifically, 
\begin{align*}
    \hat{\boldsymbol{V}}_n  =&\frac{1}{n}\sum\limits_{b=1}^n\left\{\frac{1}{\sqrt{n}}\sum\limits_{i=1}^b (\hat{\boldsymbol\theta}_i - \bar{\boldsymbol\theta}_n)\right\}\left\{\frac{1}{\sqrt{n}}\sum\limits_{i=1}^b (\hat{\boldsymbol\theta}_i - \bar{\boldsymbol\theta}_n)\right\}^\top\\
    = & \frac{1}{n^2}\left(\boldsymbol{U}_n - \bar{\boldsymbol\theta}_n \boldsymbol{v}_n^\top - \boldsymbol{v}_n\bar{\boldsymbol\theta}^\top_n + \bar{\boldsymbol\theta}_n\bar{\boldsymbol\theta}_n^\top\sum\limits_{b=1}^n b^2\right),
\end{align*}
where $\boldsymbol{U}_n = \sum_{b=1}^n\sum_{i=1}^b\hat{\boldsymbol\theta}_i\sum_{i=1}^b\hat{\boldsymbol\theta}_i^\top = \boldsymbol{U}_{n-1} + n^2\bar{\boldsymbol\theta}_n\bar{\boldsymbol\theta}_n^\top$ and $\boldsymbol{v}_n = \sum_{b=1}^n b\sum_{i=1}^b \hat{\boldsymbol\theta}_i=  \boldsymbol{v}_{n-1} + n^2\bar{\boldsymbol\theta}_n$ can both be updated recursively. Once $\bar{\boldsymbol\theta}_n$ and $\hat{\boldsymbol{V}}_n$ are obtained, the $1-\alpha_0$ asymptotic confidence interval for the $j$th element $\theta_{0,j}$ of $\boldsymbol{\theta}_0$ can be constructed as follows: 

\begin{corollary}{\label{fclt-cor2}}
    Under the conditions of Corollary \ref{fclt_cor1}, when $p$ is fixed and $n\rightarrow \infty$, we have 
    \begin{align*}
{\rm{Pr}}\left(\bar{\theta}_{n,j} - z^R_{1-\alpha_0/2}\hat{\sigma}^R_{n,j}/\sqrt{n} \leq \theta_{0,j}\leq\bar{\theta}_{n,j} + z^R_{1-\alpha_0/2}\hat{\sigma}^R_{n,j}/\sqrt{n} \right) \rightarrow 1-\alpha_0,
    \end{align*}
where $z^R_{1-\alpha_0/2}$ is ($1-\alpha_0/2$)-quantile of 
 ${\mathcal{W}_1(1)}/{[\int_0^1\{\mathcal{W}_1(r) - r\mathcal{W}_1(1)\}^2{\rm d}r ]^{1/2}}$.
\end{corollary}
The limiting distribution in (\ref{limi-dis}) is mixed normal and symmetric around zero. In practice, the critical value $z^R_{1-\alpha_0/2}$ in Corollary \ref{fclt-cor2} can be computed through Monte Carlo simulation, as detailed in Table I in \cite{abadir1997two}. 

\section{Simulation studies}\label{sec4}

In this section, we perform simulations to examine the finite-sample performance of the proposed algorithms, focusing on the coverage probabilities and lengths of confidence intervals. The methods under evaluation include the private plug-in (PPI), private batch-means (PBM), and private random scaling (PRS). As a benchmark, we also compute the non-private counterpart with plug-in (PI), batch-means (BM), and random scaling (RS). Due to space limitations, we only present the empirical performance of proposed algorithms under linear regression in our main manuscript. The results for logistic and robust expectile regressions are provided in  the Supplementary Material. 

In this simulation, we randomly generate a sequence of $n$ samples $\{(\boldsymbol{x}_n^\top,y_n)^\top\}_{n=1,2,\ldots} $ from the linear regression: \(y_n = \boldsymbol{x}_n^{\top} \boldsymbol{\theta}_0 + \varepsilon_n\), where the random noise $\varepsilon_{n}$ is i.i.d. from \(\mathcal{N}\left(0, 0.5^2\right)\) and  \(\boldsymbol{x}_n=(1, \boldsymbol{s}^\top_n)^{\top}\) denotes the covariates, with \(\boldsymbol{s}_n\) following a multivariate normal distribution \(\mathcal{N}\left(\boldsymbol{0}, \boldsymbol{\Sigma}\right)\) with two types of $\boldsymbol{\Sigma}$ structure, identity covariance $\boldsymbol{\Sigma} =\boldsymbol{I}_p$ and correlated covariance $\boldsymbol{\Sigma} = \{0.5^{|j-k|}\}_{j,k=1,\ldots,p}$. We consider a total sample size of $200,000$ arriving sequentially. For this model, the true coefficient $\boldsymbol{\theta}_0 = ({\theta}_{00},\theta_{01},\ldots,\theta_{0p})^\top= \boldsymbol{1}_{p+1}$ is a ($p+1$)-dimensional vector that includes the intercept term $\theta_{00}$. The corresponding loss function and its gradient sensitivity are outlined in Example \ref{exam1} with a step size decay rate of \( \alpha = 0.51 \). For the identity covariance matrix, we vary the parameter dimensions $p=3,5,10$ across different privacy budgets $\mu=1,2$. {A smaller privacy budget corresponds to stronger privacy protection. In addition, we set the truncation parameter \( c = 1.345 \) as used in the Huber loss and choose $M \asymp n^{0.25}$ for the batch-means estimators, as suggested by \cite{chen2020statistical}. This choice aligns with the optimal order $M \asymp n^{(1-\alpha)/2}$ when $\alpha = 1/2$. Specifically, we set $M = 20$ in our implementation. }Owing to space constraints, this section reports only the results for $p=3$. The corresponding results for $p=5$ and $p=10$  are available in   the Supplementary Material. For the correlated covariance matrix, we only present results for $p=3$, as similar patterns are observed in higher dimensions, as with the identity covariance matrix case.

The performance of each method is assessed using two criteria: the average empirical coverage
probability of the 95\% confidence interval (CP), and the average length of the 95\% confidence interval (AL). Specifically. let CI$_j$ denote a two-side 95\% confidence interval for $\theta_{0j}$, $j=1,\ldots,p$. We then calculate these two criteria across $p$ coefficients as below:
\begin{align*}
    \text{CP}=p^{-1}  \sum_{j=1}^p \widehat{\rm{Pr}}\left(\theta_{0j} \in \mathrm{CI}_{j}\right),\quad   
    \text{AL}=p^{-1} \sum_{j=1}^p \text{Length} \left(\mathrm{CI}_{j}\right),
\end{align*}
where $\widehat{\rm{Pr}}(\cdot)$ and $\text{Length}(\cdot)$ represent the empirical rate and the average length of the confidence intervals over all replications, respectively. All simulation results are based on 200 replications, with both the averages and standard errors recorded.

Figure \ref{fig:traj-linear} displays the trajectories of each component of the proposed LDP-SGD estimator $\bar{\boldsymbol{\theta}}_n$ with $\mu=1$ and $p=3$ along with three private confidence intervals for linear regression. As expected, the length of the confidence intervals decreases as $n$ increases. In the early iterations, both the estimators and confidence intervals are unstable, with noticeable fluctuations. The PRS method produces wider confidence intervals compared to PPI and PBM, indicating that PRS is more conservative than the other two methods. Additionally, the confidence interval widths for PPI and PBM are approximately equal. This observation aligns with findings from previous studies, such as those by \cite{li2022statistical} and \cite{lee2022fast}.

 We also present the empirical performance of our proposed methods under varying total sample sizes, as detailed in Table \ref{table111}. 
 A key observation is that the performance of the proposed methods aligns more closely with their non-private counterparts as the privacy budget $\mu$ increases. The coverage probabilities for all three methods approach the desired $95\%$ level as the cumulative sample size or privacy budget increases, while the lengths of the confidence intervals decrease. Among these, the PRS method exhibits more robust performance across various settings and privacy budgets, albeit with slightly wider confidence intervals than the other two. Such a wider average length could potentially account for the improvement in the average coverage rates. In contrast, the PBM method exhibits relatively lower coverage compared to the PPI and PRS methods, consistent with theoretical results indicating that PBM converges more slowly than PPI. Under privacy-preserving conditions, all three methods result in lower coverage probabilities and wider confidence intervals compared to non-private version. This effect becomes more pronounced with smaller privacy budgets, illustrating the trade-off between maintaining privacy and achieving statistical accuracy. The corresponding results for $p=5$ and $p=10$ are provided in the Supplementary Material, demonstrating similar patterns. 

 \begin{figure}[h]
  \centering
  \includegraphics[width=0.8\textwidth]{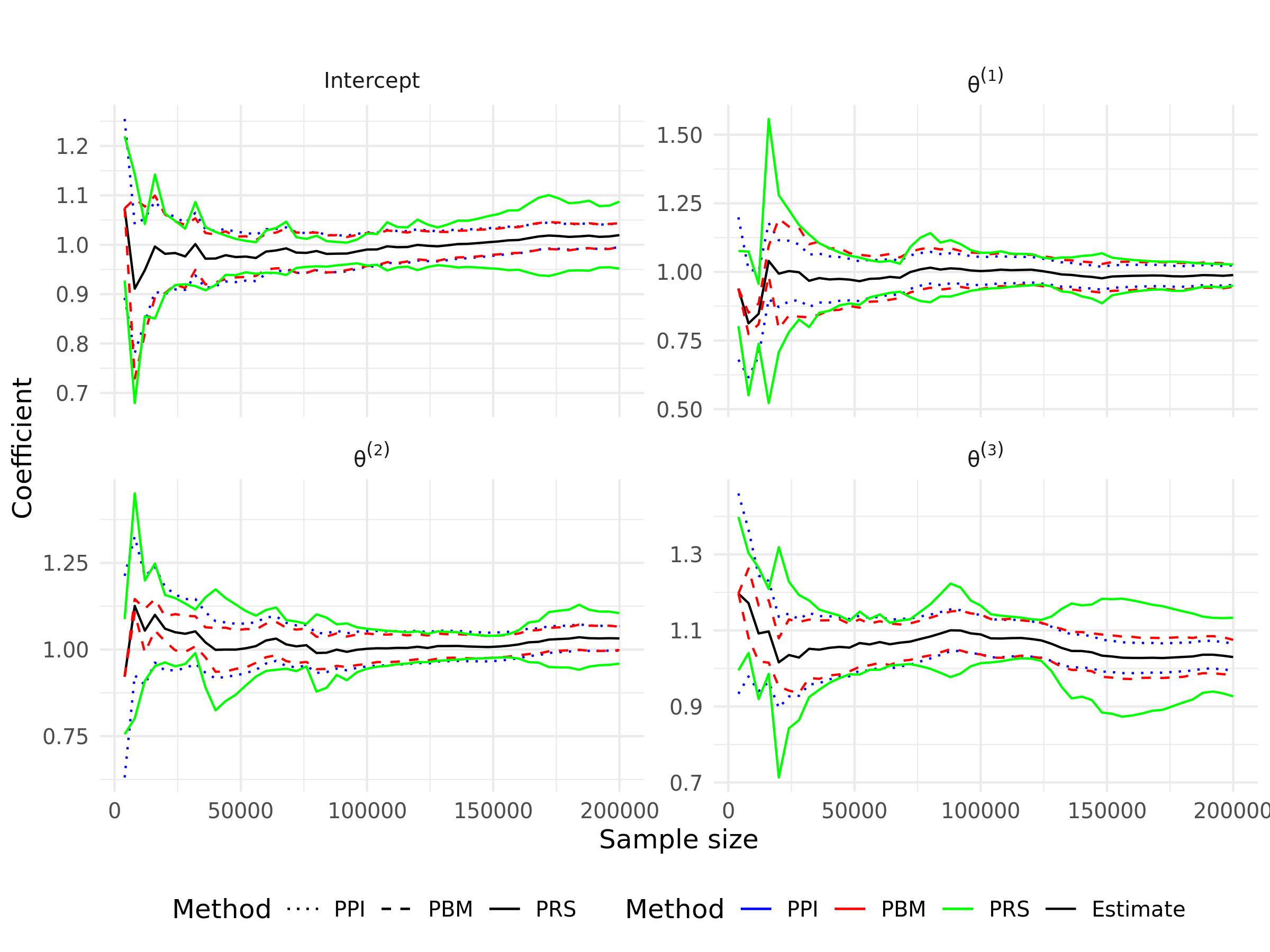}
  \caption{Trajectories of the proposed LDP-SGD estimator $\bar{\boldsymbol{\theta}}_n$ (black solid line) along with three corresponding private confidence intervals, for linear regression with $p=3$ and $\mu=1$, where $\theta^{(i)}$ represents the $i$-th coordinate, plotted against cumulative sample size.} 
  \label{fig:traj-linear}
\end{figure}

\begin{table}
\centering
\caption{Linear regression: Summary of averages and standard errors (brackets) of the coverage probabilities ($\%$) and length of the 95\% confidence interval ($\times 10^{-2}$) across various scenarios with $p=3$. CP:  the coverage probability;
AL: the average length of the 95\% confidence interval.}
{
\fontsize{9.5}{9.5}\selectfont
\label{table111}
\begin{tabular}{lrrrrr}
\hline 
 & $n=40000$ & $n=80000$ & $n=120000$ & $n=160000$ & $n=200000$ \\
\hline 
\multicolumn{6}{c}{\qquad\qquad Non-private}\\
\text{PI: CP} & 93.75 (3.66) & 92.50 (4.83) & 93.50 (3.54) & 92.75 (3.97) & 94.13 (3.09)\\
\text{PI: AL} & 1.08 (0.20) & 0.76 (0.14) & 0.62 (0.11) & 0.54 (0.10) & 0.48 (0.09)\\
\text{BM: CP} & 88.88 (2.01) & 90.75 (1.55) & 93.25 (2.50) & 93.13 (1.55)  & 92.75 (1.26)\\
\text{BM: AL} & 1.01 (0.02) & 0.74 (0.01) & 0.62 (0.01) & 0.54 (0.01) & 0.48 (0.01)\\
\text{RS: CP} & 95.13 (1.38) &  94.00 (1.47) & 95.75 (0.50) & 95.13 (0.75) & 95.50 (1.22)\\
\text{RS: AL} & 1.47 (0.02) & 1.03 (0.02) & 0.84 (0.02) &  0.73 (0.03) & 0.64 (0.02)\\
\multicolumn{6}{c}{\qquad\qquad $\mu=1$}\\
\text{PPI: CP} & 91.38 (3.82) & 93.50 (4.30) & 92.13 (5.34)  & 93.75 (3.18) & 93.25 (4.50)\\
\text{PPI: AL} & 11.49 (0.38) & 7.68 (0.19) & 6.10 (0.13) & 5.19 (0.10) & 4.60 (0.07)\\
\text{PBM: CP} & 85.38 (1.60) & 92.00 (1.41) & 91.63 (2.93) & 93.25 (2.02) & 92.88 (2.93)\\
\text{PBM: AL} & 11.09 (2.28) & 7.83 (1.57) & 6.49 (1.29) & 5.49 (1.05) & 4.77 (0.90)\\
\text{PRS: CP} & 94.38 (1.93) & 95.75 (0.29) & 94.88 (1.44) & 95.38 (2.32) & 95.50 (1.08)\\
\text{PRS: AL} & 16.01 (3.31) & 10.96 (2.21) & 8.64 (1.78) & 7.21 (1.28) & 6.50 (1.15)\\
\multicolumn{6}{c}{\qquad\qquad $\mu=2$}\\
\text{PPI: CP} & 90.75 (6.25) & 90.50 (4.88) & 91.25 (4.29) & 92.63 (3.42) & 92.13 (4.27)\\
\text{PPI: AL} & 4.99 (0.07) & 3.45 (0.08) & 2.81 (0.10) & 2.42 (0.10) & 2.16 (0.10) \\
\text{PBM: CP} & 84.63 (2.78) & 87.88 (2.63) & 89.75 (1.31) & 90.25 (1.00) & 90.88 (2.10)\\
\text{PBM: AL} & 4.70 (0.08) & 3.42 (0.06) & 2.87 (0.05) & 2.47 (0.04) & 2.15 (0.04)\\
\text{PRS: CP} & 92.75 (3.50) & 92.50 (1.78) & 93.50 (1.58) & 94.00 (1.62) & 94.88 (1.61)\\
\text{PRS: AL} & 6.77 (0.11) & 4.82 (0.09) & 3.94 (0.08) & 3.30 (0.06) & 2.93 (0.05)\\
\hline
\end{tabular}
}

\end{table}

{
To further evaluate the coverage properties of PPI and PRS, we vary the significance level $\alpha_0$ and compute the corresponding $1 - \alpha_0$ asymptotic confidence intervals, alongside the empirical coverage probabilities. Specifically, $\alpha_0$ is varied from $0.01$ to $0.1$ in increments of $0.01$. The analysis is carried out for a linear regression model with dimension $p=5$ and sample size $n=200000$, under two different privacy budgets $\mu=1,2$. The results are presented in Figure \ref{fig:cp_vs_alpha}. For $\mu=1$, the PRS method exhibits a more conservative coverage probability, slightly surpassing the oracle coverage level, indicating minor over-coverage. In contrast, PPI tends to under-cover with its narrower confidence intervals. As the privacy budget increases to $\mu=2$, both PRS and PPI coverage probabilities shift closer to the oracle curve. 

}

\begin{figure}
  \centering
  \includegraphics[width=8.0cm,height=8.0cm]{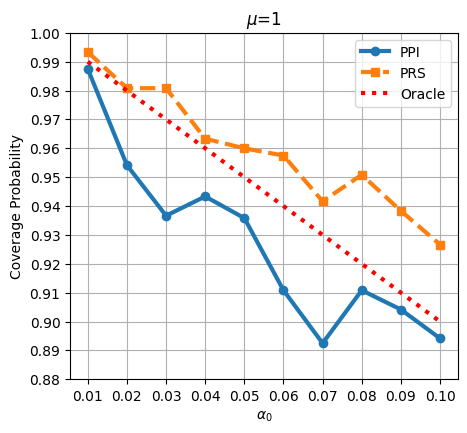}\includegraphics[width=8.0cm,height=8.0cm]{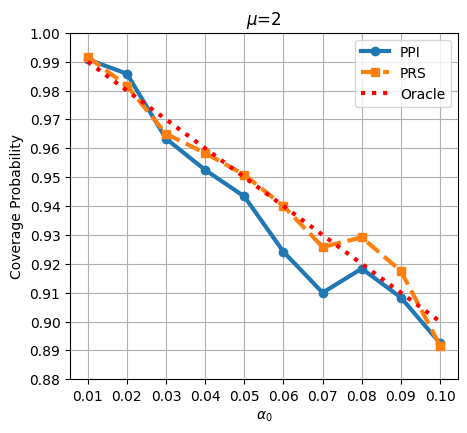}
  \caption{Coverage probability versus nominal level $\alpha_0$ for PPI and PRS with dimension $p=5$ and sample size $n=200000$ under different privacy budget.} 
  \label{fig:cp_vs_alpha}
\end{figure}

\section{Applications}\label{sec5}

{
In this section, we illustrate the effectiveness of the proposed method by analyzing two real-world datasets: the Ride-sharing data and the US insurance data.

\subsection{Ride-sharing Data}
In this subsection, we apply the proposed LDP-SGD algorithm to the Ride-sharing dataset, which contains synthetic data from a ride-sharing platform. The dataset spans from January 2024 to October 2024 and includes detailed records of rides, users, drivers, vehicles, and ratings. It offers valuable insights into driver performance and fare prediction over time. The dataset is publicly available at \textsf{https://www.kaggle.com/datasets/adnananam/ride-sharing-platform-data}, comprising 50,000 ride records and 3,000 profiles of drivers and vehicles. For preprocessing, we concatenate driver and vehicle information with each corresponding ride. We then select the following features as potential predictors for fare estimation: ride distance, driver rating, total number of rides completed by the driver, vehicle production year, vehicle capacity, and ride duration. After concatenation, the dataset is standardized by subtracting the mean and dividing by the standard deviation for each column. The resulting data is then used as input to the LDP-SGD algorithm with weighted Huber loss defined in Example 1, which is run with a privacy budget of $\mu = 1$ and a step size decay rate of $\alpha = 0.501$. 

\begin{figure}[h]
  \centering
  \includegraphics[width=0.8\textwidth]{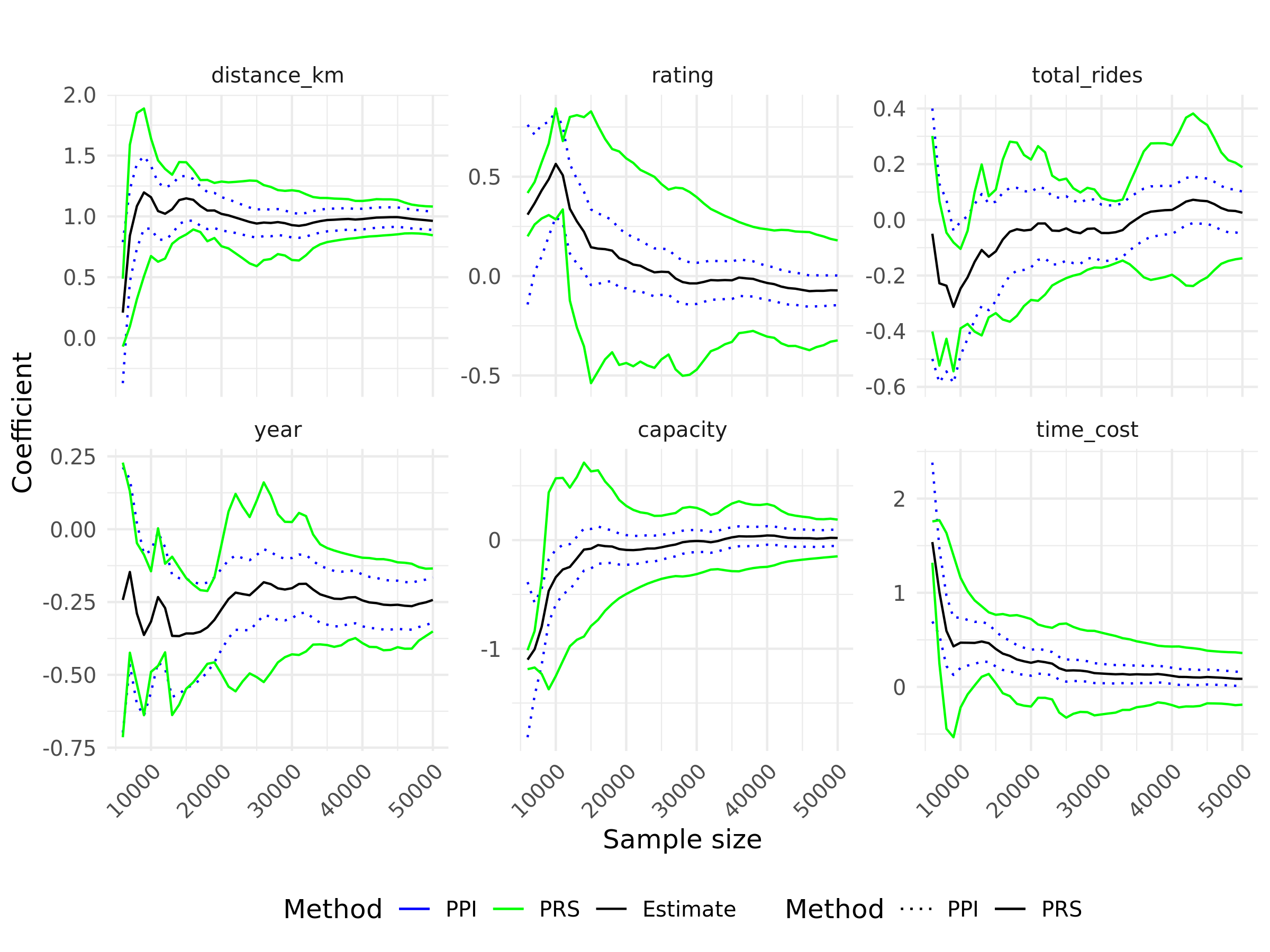}
  \caption{Trajectories of the LDP-SGD estimators (black solid line) in the Ride-sharing dataset predicting fare amount, accompanied by two private confidence intervals with $\mu=1$.} 
  \label{fig:rideshare}
\end{figure}

The results of this analysis are presented in Figure \ref{fig:rideshare}. As the sample size increases, the coefficients gradually stabilize. Specifically, the coefficients for 'distance' and 'time cost' converge to positive values, with both the PPI and PRS confidence intervals entirely above zero. This suggests a significant positive impact of these factors on fare amount, aligning with intuitive expectations, as longer distances and extended journey times are naturally associated with higher costs. In contrast, the coefficients for 'rating' and 'total rides' converge toward values near zero, with their confidence intervals including zero, indicating that driver-related attributes do not meaningfully affect fare pricing. Similarly, the coefficients for 'year' and 'capacity' also converge to values close to zero, with their confidence intervals encompassing zero, suggesting that vehicle-related factors have no significant influence on the fare amount. Overall, the analysis shows that only distance and time cost significantly impact fare pricing, while driver and vehicle characteristics have negligible effects. This is likely because ride fares are primarily driven by distance, duration, and demand fluctuations, which outweigh the influence of individual driver and vehicle attributes.

\subsection{US insurance Data}\label{sec:reta}

In this subsection, we conduct a analysis on the Insurance dataset, which aims to predict health insurance premiums in the United States. The dataset, based on synthetic data, captures a variety of factors that influence medical costs and premiums. It includes the following variables on the insurance customers: age, gender, body mass index (BMI), number of children, smoking status, region, medical history, exercise frequency, occupation, and type of insurance plan. The full list of variables is presented in Table \ref{tab:variables}. This dataset was generated using a script that randomly sampled 1,000,000 records, ensuring a comprehensive representation of the insured population in the US. Publicly available at \textsf{https://www.kaggle.com/datasets/sridharstreaks/insurance-data-for-machine-learning/data}, it provides valuable insights into the factors affecting health insurance premiums.

\begin{table}[ht]
\centering
\fontsize{9.5}{9.5}\selectfont
\begin{tabular}{|l|l|l|}
\hline
\textbf{Variable Name} & \textbf{Type} & \textbf{Range / Categories} \\
\hline
age & Numerical & $[18, 65]$ \\
gender & Categorical (Binary) & \{Female, Male\} \\
bmi & Numerical & $[18, 50]$ \\
children & Numerical (Integer) & $\{0,1,2,3,4,5\}$ \\
smoker & Categorical (Binary) & \{No, Yes\} \\
medical\_history & Categorical (Ordinal) & \{None, blood pressure, Diabetes, Heart Disease\} \\
family\_medical\_history & Categorical (Ordinal) & \{None, blood pressure, Diabetes, Heart Disease\} \\
exercise\_frequency & Categorical (Ordinal) & \{Never, Occasionally, Rarely, Frequently\} \\
occupation & Categorical (Ordinal) & \{Student, Unemployed, Blue collar, White collar\} \\
coverage\_level & Categorical (Ordinal) & \{Basic, Standard, Premium\} \\
charges & Numerical & $[3445.01, 32561.56]$ \\
\hline
\end{tabular}
\caption{Summary of variables for US insurance data with their types and ranges}
\label{tab:variables}
\end{table}

For the analysis, we first carried out extensive data preprocessing to ensure proper handling of both numerical and categorical variables. Numerical variables, such as age, BMI, and the number of children, were standardized to have a mean of zero and a standard deviation of one, ensuring consistency across features for the modeling process. Categorical variables were encoded with meaningful numeric values to reflect their underlying semantics. Specifically, health-related variables like \textit{medical\_history} and \textit{family\_medical\_history} were assigned values based on the severity of conditions: ``None'' (0), ``High blood pressure'' (1), ``Diabetes'' (2), and ``Heart disease'' (3). The \textit{exercise\_frequency} variable was encoded as ``Never'' (0), ``Occasionally'' (1), ``Rarely'' (2), and ``Frequently'' (3), reflecting varying levels of activity. For \textit{occupation}, a hierarchical encoding was used, ranging from ``Student'' (0) to ``White collar'' (3), capturing the occupational spectrum. Similarly, the \textit{coverage\_level} variable was encoded as ``Basic'' (0), ``Standard'' (1), and ``Premium'' (2) to reflect the varying levels of insurance coverage. Binary variables such as \textit{smoker} and \textit{gender} were encoded as 0 and 1, representing ``no''/``yes'' or ``female''/``male,'' respectively. The processed dataset was randomly split into training and testing sets, consisting of 800,000 and 200,000 observations, respectively. The training set is then input into the LDP-SGD algorithm with a weighted Huber loss, using a privacy budget of $\mu = 1$ and a step size decay rate of $\alpha = 0.501$. The corresponding results observed during the training process are presented in Figure~\ref{fig:insurance}.

\begin{figure}[h]
  \centering
  \includegraphics[width=0.9\textwidth]{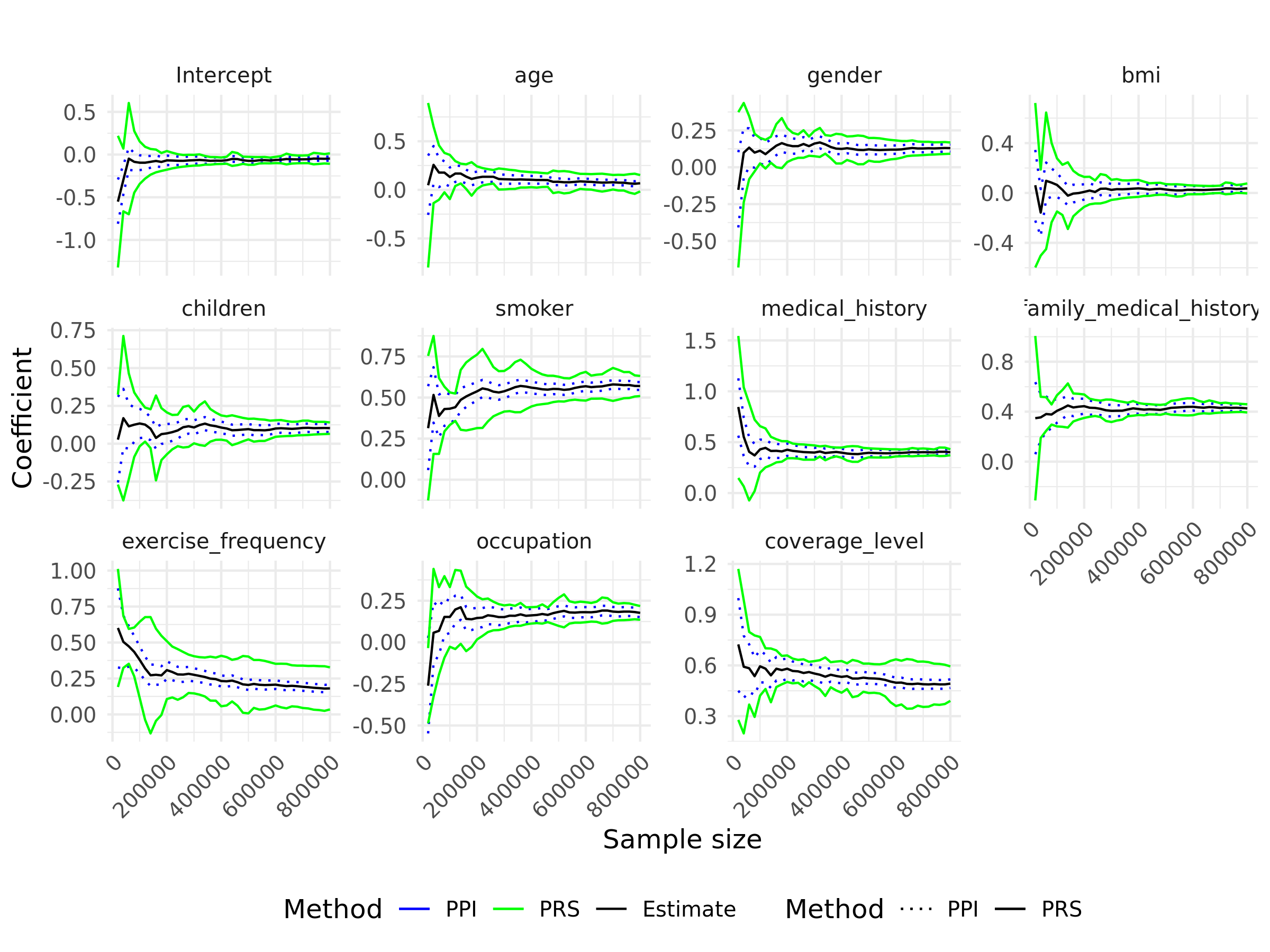}
  \caption{Trajectories of the LDP-SGD estimators (black solid line) for each variable in the US insurance dataset, along with two corresponding private confidence intervals with $\mu=1$.} 
  \label{fig:insurance}
\end{figure}

The results demonstrate that as the sample size increases, all regression coefficients for predicting health insurance premiums converge to positive values, indicating their significant contributions to premium determination. A positive coefficient implies that higher values of a predictor correspond to higher premiums. For instance, males are charged higher premiums than females, smokers incur substantially higher costs due to increased health risks, and individuals with serious medical conditions or a family history of such conditions (e.g., diabetes, heart disease) face elevated premiums. Likewise, a higher BMI and selecting more comprehensive insurance plans also lead to higher charges, reflecting the associated health risks or benefits. While the coefficients for age and exercise frequency are positive, their magnitudes are relatively small, suggesting a limited linear effect on premium pricing. Its influence may be non-linear and not fully captured by a simple linear term. Overall, the analysis highlights that risk-related factors such as smoking, BMI, and medical history, are the primary drivers of health insurance premiums, while demographic and lifestyle factors exert a lesser influence.

We apply the LDP-SGD estimator to perform linear regression for predicting insurance charges. The mean squared error (MSE) on the test set is 0.0722. For comparison, we also fit an offline ordinary least squares (OLS) estimator without privacy guarantees using the same training data, which achieves an MSE of 0.0607.
Table~\ref{tab:estimation_comparison} summarizes the comparison between the LDP-SGD and offline OLS estimators in terms of coefficient estimates. The results show that LDP-SGD estimators closely align with OLS estimates, indicating that the proposed algorithm effectively protects privacy while preserving utility and explanatory power.  

\begin{table}[h]
    \centering
    \renewcommand{\arraystretch}{0.5}
    \begin{tabular}{lccccc}
        \toprule
        Setting & age & gender & bmi & children & smoker \\
        \midrule
        OLS & 0.0627 & 0.1131 & 0.1044 & 0.0774 & 0.5659 \\
        LDP-SGD  & 0.0679 & 0.1298 & 0.0664 & 0.0828 & 0.5483 \\
        \midrule
        Setting & medical & family medical & exercise & occupation & coverage level \\
        \midrule
        OLS & 0.4052 & 0.4051 & 0.1389 & 0.1516 & 0.4625 \\
        LDP-SGD  & 0.4009 & 0.4264 & 0.1409 & 0.1767 & 0.4634 \\
        \bottomrule
    \end{tabular}
    \caption{Comparison of offline OLS and online LDP-SGD estimations.}
    \label{tab:estimation_comparison}
\end{table}

}



 \section{Discussion} \label{sec:conclusion}
 
This paper investigates online private estimation and inference for optimization-based problems with streaming data under local differential privacy constraints. To eliminate the dependence on trusted data collectors, we propose an LDP-SGD algorithm that processes data in a single pass, making it well-suited for streaming applications and minimizing storage costs. Additionally, we introduce and discuss three asymptotically valid online private inference methods—private plug-in, private batch-means, and random scaling—for constructing private confidence intervals. We establish the consistency and asymptotic normality of the proposed estimators, providing a theoretical foundation for our online private inference approach.

Several directions for future research are worth exploring. First, while this work focuses on low-dimensional settings, extending the methodology to high-dimensional penalized estimators using noisy proximal methods would be valuable. Second, our models are based on parametric assumptions; adapting these to non-parametric models by developing a noisy functional SGD algorithm remains interesting. {Finally, this work assumes strong convexity and smoothness of the objective loss function. However, extending the proposed framework to encompass general nonsmooth convex functions, such as ReLU and quantile losses, which are piecewise linear and lack strong convexity, call for future research.}

\bibliography{reference}
\bibliographystyle{chicago}

\end{document}